# A change of origin and an update to SIS100 simulation geometries

*E. Clerkin* *[1] *and P. Dahm* [2]
[1]FAIR, Darmstadt, Germany; [2]GSI, Darmstadt, Germany

## 1. Introduction

For last year's progress report, the two authors had conducted a survey of differences between technical and simulation geometries, and published in Ref. [1], the recommended detector positions of the simulation (GEANT) geometries to best align with the most recent technical computer aided designs (CAD). In that 2020 document, it is strongly stressed that detector placements were likely to change in the course of the coming year, in response to the allowances for support structure and improved technical designs and in the main, this is exactly what occurred.

One such substantive change was the anticipated elongation of the box enclosure of silicon-tracking-system detector (STS), warranted by an internal expansion of the spacing between sensor layers. It had been believed to make space of the box enlargement, the target would need to shift upstream by 4 cm, in order to avoid collisions with other volumes, and would thus affect many detector subsystems by increasing their distance to the target. It had been decided by the technical board in April 2020, that such a change request could be accepted but only once a new STS simulation geometry, which realised these changes in pitch and length was developed and made available to the community. These STS geometries were developed by Mehul Shiroya during the course of the year, and was distributed to the wider collaboration in November 2021 via its inclusion in the official geometry repository available at https://git.cbm.gsi.de/CbmSoft/cbmroot_geometry. Its development is documented in Ref. [2] of this report. It should be perhaps noted that after full implementation of this change, and due to small miscellaneous changes which saved some space, the target shift may not have been needed as the STS still would have a 1.2 cm clearance to neighbouring matter. It is therefore possible that this decision may be revisited in the near future, especially if simulations show detrimental performance of our detectors. Major geometry changes during the year 2021 with a particular focus on changes that occurred across CBM subgroups will be discussed in Sec. 3 of this manuscript which addresses the default experimental setups which coincide with our DEC21 software release.

In addition to these many changes, it was decided to act now on the widely held desire to change the origin of the CBM experiment. Since its inception, the origin had been naturally defined as the target position, which has a conceptually convenient location to imagine the initial beam particle interactions. However, it suffers from a lack of fixed-position status which is believed to be crucial for direct comparison between the CAD and the GEANT geometries. This was partly emphasised by the STS change which moved the target, meaning that the origin would need to be defined as a distance along the nominal beam axis from another fixed point. Additionally recent requests have been made that the target should be capable of being moved depending on the beam momentum. The design of target exchange device, Ref [3], proposed during the course of the year hasn't this capability yet. The magnet is a massive fixed structure, centred symmetrically around the nominal beam axis. Its centre of the mass, coinciding with a non-physical point in space, under vacuum, inside the beampipe, inside the STS box, is a natural reference point defined by the magnetic field of our experiment. The nominal beam axis and the vertical direction through the centre of the magnet provides sufficient orientation to this point, which, along with standard distance measures, allows us to define a coordinate system which satisfies our technical requirements whilst being non-arbitrarily defined. It is a special point being the centre of the magnetic field. It was thus decided that the *centre of the magnet* should become our new origin. Although such a change would and did erect many challenges for development of our software framework (CBMROOT), it was expressed by senior developers at our software meetings, that changing the origin within the CBMROOT simulation environment, may even help find issues within our code base which may, in the long-run, lead to a more robust software environment. In Sec. [2] of this report, this information is documented for perpetuity and recorded for future reference by the collaboration.

When also considering that the CBMROOT code base changed considerably between the APR21 release and the DEC21 software release with the additions of several new features, coordination of the realisation of the last two paragraphs into our simulation geometries introduces a real dilemma on how to maintain control when making two or more changes to each detector geometry. In order to test and validate a software release with new features, a nominal change in the position of the target, a new defining origin and coordinate system, plus the normal incremental improvements to our subsystem geometries, a two step approach was pursued. In the next section of this report, we updated all default geometries in Ref. [1] from our APR21 software release with the newly defined origin. In the subsequent section, we discuss new geometries introduced for DEC21 with particular emphasis on our new DEC21 geometry defaults.

---

*e.clerkin@gsi.de

## 2. Coordinate Change (APR21 → APR21+)

Several detectors had their geometries updated which often changed their length as well as their placements. New space provisions needed to be accounted for, in particular, for incorporation of the proposed beam-pipe designs, cf. Refs. [4, 5] of this report. A target shift was planned and a coordinate system which redefined the origin from the target position to the centre of the magnet. A script which switched between the two coordinate systems, and the candidate default geometries was distributed in the macro geometry folder of CBMROOT and was announced at the CBMCM (Ref. [6]) Uptake of this tool by the community was however disappointing, probably due to a lack of manpower across CBM subgroups. Once the feature freeze for the DEC21 software release was announced, a two stage approach to transition from the APR21 default geometries was in the end forced so that consequence of each change on the software could be traced effectively. A new intermediate geometry for each subsystem was created to be identical to the APR21 default geometry but with the new coordinate system. For most geometries, this meant a defacto translation of the geometry upstream by 40 cm to account for the centre of the magnet being 40 cm downstream from the old target position. For geometries such as the STS, RICH, MUCH and PIPE the translation occurred in the near top volumes of the geometries whilst the TRD, TOF, and PSD had their positional matrices directly modified. Their macros were supplied to CBMROOT and corresponding binaries were made available via the geometry repository. Additionally new parameter and material budget files also needed to be created in the parameter repository, available from https://git.cbm.gsi.de/CbmSoft/cbmroot_parameter. The same MVD as in Ref. [7], could be used without incorporation of the coordinate change as it is essentially embedded into a volume, named "pipevac1", of the beam-pipe which itself moved. However, during the testing process, MVD geometries v20c, v20d were requested by the tracking team for use with the beam-pipe which did not shift the "pipevac1" volume. One of these beampipe geometries (Ref. [2]) is the current default so these MVD geometries are also used as defaults. Simulation versions of complete technically feasible candidate beampipe, Refs [8, 9], are distributed in the DEC21 release although not as defaults as no testing period could be performed.

These geometries which could be directly compared to the APR21 defaults but defined in new coordinate system of the magnetic field were colloquially called APR21+ setups where the plus symbol refereed to the addition of a new origin. For naming of the APR21+ geometries, an increment of the letter index of the tag rather than an update of the number which represents the year was preferred so as to not be confused with geometries with new features. The RICH detector and PLATFORM geometry were exceptions to this loose rule. The APR21+ geometries are listed in Tab. 1 and may be compared with the APR21 geometries in Tab. [1] of Ref. [1]. The testing period for these geometries turned out to be extensive as this change in particular revealed several issues in CBMROOT after the DEC21 code freeze. Namely changes to TOF and RICH tracking needed to be fixed before the final release. It is expected that the APR21+ geometries, not used as DEC21 defaults, will be removed from distribution in the near future.

## 3. New Defaults (APR21+ → DEC21)

In relation to CAD updates during the past year, many of the systems, that will make up the CBM Experiment moved closer to production. As such, the level of detail in the planning and CAD modelling increased significantly. Major changes include technical geometries for MVD, STS, MUCH, TRD, TOF and PSD, which were all submitted in the time after last years progress report. Apart from detectors, a lot of infrastructure was included in the models, as shown in the collaboration meeting in late 2021. Very recently, the Cryo installation for the magnet and the steel construction for the upstream platform were updated with close to final designs and a concept for the beampipe was included. Since each of these points is described in more detail in the respective subsystem sections of this report, only a summary of the currently used models and their submission dates is given in the following list. Current used models in CAD were last updated in:

- MVD 07.04.2021
- STS 20.02.2022
- RICH 04.11.2020
- MUCH 29.09.2021
- TRD 08.10.2021
- TOF 15.02.2022
- PSD 08.04.2021
- Magnet 25.06.2021
- Cryo 08.12.2021
- Beampipe 21.02.2022
- Platform 03.02.2022

| SUB | Electron | Hadron | J/$\psi$ | LMVM |
|---|---|---|---|---|
| MAG | v21a | v21a | v21b | v21b |
| PIPE | v16e_1e | v16e_1e | v20b_1m | v20b_1m |
| MVD | v20a_tr | - | - | - |
| STS | v20a | v20a | v20a | v20a |
| RICH | v21a | - | - | - |
| MUCH | - | - | v20c.*jpsi | v20c.*lmvm |
| TRD | v20c_1e | v20c_1h | v20c_1m | v20c_1m |
| TOF | v20c_1e | v20c_1h | v20c_1m | v20c_1m |
| PSD | v20c | v22c | - | - |
| PLAT | v13a | v13a | v13a | v13a |

Table 1: APR21+ geometries. Geometries required in a validation and testing stage for the DEC21 SW release. Simulation of electron, hadron, J$\psi$ and LMVM experimental setups of the CBM experiment at FAIR. The '*' abbreviates '_sis100_1m_' in the tag name.

| SUB | Electron | Hadron | J/$\psi$ | LMVM |
|---|---|---|---|---|
| MAG | v21a | v21a | v21b | v21b |
| PIPE | v21d | v21d | v21d | v21d |
| MVD | v20d_tr | - | - | - |
| STS | v21e | v21e | v21e | v21e |
| RICH | v21a | - | - | - |
| MUCH | - | - | v21c*jpsi | v21c*lmvm |
| TRD | v20b_1e | v20c_1h | v20c_1m | v20c_1m |
| TOF | v21a_1e | v21a_1h | v21a_1m | v21a_1m |
| PSD | v22a | v22c | - | - |
| PLAT | v22b | v22b | v22b | v22b |

Table 2: DEC21 geometries. The 2021 geometry versions in the DEC21 software release for simulation of electron, hadron, J$\psi$ and low mass vector meson (LMVM) experimental setups of the CBM experiment at FAIR. The '*' abbreviates '_sis100_1m_' in the tag name.

The rest of this section will address the effect of convergence to new technical redesigns on the simulation geometries. Several geometries with new layouts were introduced in the DEC21 release, and so were not directly comparable to their APR21 counterparts. In many cases, an origin-changed geometry also needed to be created, along with the binaries and parameter in the same fashion as was done in previous section. This occurred due to the geometry being created before the origin change was demanded at the geometry submission stage. For example, the MUCH geometries (Ref. [10]) were already made available before the APR21 software release, but were not included at that time as default geometries due to insufficient testing and validation. New MUCH geometries were created for the release.

It was decision in the SWM in February 17th 2022 that the PSD should be placed away from the beam in its parking position, v22c, for the hadron setup. Tab. 2 contains a list of the current setups and the default geometries to be used for each setup. Since APR21 release (Ref. [1]), many of the geometries were updated, upgraded, and redesigned.

Table 2 contains a list of the geometries in the DEC21 release and Tab. 3 contains the positions of all detector subsystems from the centre of the magnetic origin point. Some additional changes of the distance between the target and detector as compared with Tab. [2] and [3] of Ref. [1]. It is noted that space provision for inclusion of a bellows assembly (Ref. [5]) matched the displacements between the old target position and the centre of the magnet thereby making geometries appear to maintain their positions whilst in fact being 44 cm further from the nominal target position. Also as a consequence, there is now no significant distinction of the possible positioning of downstream detectors between the muon and electron setups.

Unlike in Ref. [1], provision of the beam-fragmentation T0 counter is, at the very minimum, paused due to Russia's invasion of the Ukraine which has lead to Western countries to sanction Russia, which has had huge implementation for international scientific facilities and collaborations such as FAIR and CBM. Changes in technical and simulation geometries, and the experimental setups will likely change dramatically as an indirect consequence.

| SUB | width | height | length | position |
|---|---|---|---|---|
| MAG | 4.4 | 3.7 | 2.0 | -0.2 |
| MAG (clamps) | | | 2.38 | -0.39 |
| MVD (TR) | 0.8 | 0.8 | 0.525 | -0.36 |
| MVD (VX) | | | | -0.39 |
| STS | 2.38 | 1.423 | 1.338 | -0.1 |
| RICH | 6.0 | 5.478 | 2.2 | 1.388 |
| MUCH (J/$\psi$) | 5.0 | 5.0 | 3.52 | 0.85 |
| MUCH (LMVM) | 3.88 | 4.03 | 2.5 | 0.85 |
| 5th Absorber | 5.0 | 5.0 | 1.0 | 3.27 |
| TRD | 9.9 | 8.55 | 2.9 | 4.4 |
| TOF | 13.5 | 10.78 | 2.19 | 6.9 |
| PSD | 4.85 | 10.04 | 1.96 | 10.1 |
| lat. | | | | -0.9 |
| vert. | | | | 0.7 |
| angle | | | | 3° |

Table 3: Dimensions of detectors and position distances from the origin defined as the centre of the magnet. All measurements are in meters, width is the left to right view looking at the subsystem (SUB) from the beam perspective, height is vertical, and lengths are along the beam axis. The subsystems are centred on the beam axis.